\documentclass[11pt]{article}
\usepackage{epic,eepic}
\usepackage{amsmath,amssymb,amsfonts}
\begin{document}
\newtheorem{theorem}{Theorem}[section]
\newtheorem{corollary}[theorem]{Corollary}
\newtheorem{lemma}[theorem]{Lemma}
\newtheorem{remark}[theorem]{Remark}
\newtheorem{proposition}[theorem]{Proposition}
\newtheorem{definition}[theorem]{Definition}
\def\emptyset{\varnothing}
\def\setminus{\smallsetminus}
\def\Irr{{\mathrm{Irr}}}
\def\Rep{{\mathrm{Rep}}}
\def\End{{\mathrm{End}}}
\def\Tube{{\mathrm{Tube}}}
\def\Vec{{\mathrm{Vec}}}
\def\loc{{\mathrm{loc}}}
\def\opp{{\mathrm{opp}}}
\def\id{{\mathrm{id}}}
\def\A{{\mathcal{A}}}
\def\B{{\mathcal{B}}}
\def\C{{\mathcal{C}}}
\def\D{{\mathcal{D}}}
\def\E{{\mathcal{E}}}
\def\X{{\mathcal{X}}}
\def\CC{{\mathbb{C}}}
\def\N{{\mathbb{N}}}
\def\Q{{\mathbb{Q}}}
\def\R{{\mathbb{R}}}
\def\Z{{\mathbb{Z}}}
\def\a{{\alpha}}
\def\be{{\beta}}
\def\de{{\delta}}
\def\e{{\varepsilon}}
\def\si{{\sigma}}
\def\la{{\lambda}}
\def\th{{\theta}}
\def\lan{{\langle}}
\def\ran{{\rangle}}
\def\isom{{\cong}}
\newcommand{\Hom}{\mathop{\mathrm{Hom}}\nolimits}
\def\qed{{\unskip\nobreak\hfil\penalty50
\hskip2em\hbox{}\nobreak\hfil$\square$
\parfillskip=0pt \finalhyphendemerits=0\par}\medskip}
\def\proof{\trivlist \item[\hskip \labelsep{\bf Proof.\ }]}
\def\endproof{\null\hfill\qed\endtrivlist\noindent}

\title{A remark on matrix product operator algebras, \\
anyons and subfactors}
\author{
{\sc Yasuyuki Kawahigashi}\\
{\small Graduate School of Mathematical Sciences}\\
{\small The University of Tokyo, Komaba, Tokyo, 153-8914, Japan}
\\[0,40cm]
{\small Kavli IPMU (WPI), the University of Tokyo}\\
{\small 5-1-5 Kashiwanoha, Kashiwa, 277-8583, Japan}
\\[0,05cm]
{\small and}
\\[0,05cm]
{\small iTHEMS Research Group, RIKEN}\\
{\small 2-1 Hirosawa, Wako, Saitama 351-0198,Japan}\\
{\small e-mail: {\tt yasuyuki@ms.u-tokyo.ac.jp}}}
\maketitle{}
\begin{abstract}
We show that the mathematical structures in a recent work of
Bultinck-Mari\"ena-Williamson-\c Sahino\u glu-Haegemana-Verstraete
are the same as those of flat symmetric bi-unitary
connections and the tube algebra in subfactor theory.
More specifically, a system of flat symmetric
bi-unitary connections arising from a subfactor with finite
index and finite depth
satisfies all their requirements for tensors and the tube algebra
for such a subfactor and the anyon algebra for such tensors are
isomorphic up to the normalization constants.  Furthermore,
the matrix product operator algebras arising from
tensors corresponding to possibly non-flat symmetric
bi-unitary connections are isomorphic to those
arising from flat symmetric bi-unitary connections
for subfactors.
\end{abstract}

\section{Introduction}

A recent work
Bultinck-Mari\"ena-Williamson-\c Sahino\u glu-Haegemana-Verstraete
\cite{BMWSHV} has caught much attention in the community of
condensed matter physics in connection to 2-dimensional topological
phases of matter.  
(See \cite{KLPG}, \cite{YGW}, for example.  They are also related
to the Levin-Wen model \cite{LW}.)
As they themselves are aware and give a
citation to \cite{EK2}, the mathematical
structures in their work are similar to those in subfactor theory.
In this short note, we present precise mathematical relations 
between the two settings.  In fact, under some natural setting, we show
that the machinery in \cite{BMWSHV} is mathematically the same
as that of flat symmetric bi-unitary connections of Ocneanu \cite{O1},
\cite[Chapter 10]{EK3}.  We also
discuss how to deal with non-flat symmetric
bi-unitary connections which
also naturally appear in this framework.

The Jones theory of subfactors \cite{J1} opened a vast new field
in theory of operator algebras and his discovery of the Jones
polynomial \cite{J2} initiated huge interest in 
\textsl{quantum} invariants in 3-dimensional
topology.  New algebraic
structures governing such mathematics are fusion categories
and modular tensor categories.  See \cite{K2} and references
therein for operator algebraic approaches to these categories
and connections to 2-dimensional conformal field theory.
A modular tensor category is a useful tool
to study \textsl{anyons} in 2-dimensional phases of matter.
An anyon is a quasi-particle in 2-dimensional phases of matter
and its exchanges follow braid group statistics.
This viewpoint is also expected to be useful for mathematical
understanding of topological quantum computations \cite{W}.
We would like to present how to use operator algebraic
tools to study mathematical problems on anyons related to
\cite{BMWSHV}.

The author thanks Microsoft Station Q for
hospitality where a part of this work was done.
The author gratefully acknowledges support from the Simons Center
for Geometry and Physics, Stony Brook University at which some
of the research for this paper was performed.
This work was also partially supported by 
JST CREST program JPMJCR18T6 and
Grants-in-Aid for Scientific Research 19H00640.  
The author thanks N. Bultinck for explaining the contents
of \cite{BMWSHV} in three lectures at Tokyo and
Y. Ogata for calling attention to paper
\cite{BMWSHV} and helpful discussions.

\section{Symmetric bi-unitary connections}

In order to compare the framework of a recent paper
Bultinck-Mari\"ena-Williamson-\c Sahino\u glu-Haegemana-Verstraete
\cite{BMWSHV} with that in subfactor theory, we need to
present machinery of bi-unitary connections \cite[Chapter 11]{EK3}
in a little bit different setting from that in \cite{EK3} so
that we can establish a direct relation to that of
tensors in \cite{BMWSHV}.
In this paper, a subfactor always means a subfactor
of the hyperfinite II$_1$ factor with finite Jones index
and finite depth.  We refer the reader to \cite[Chapter 9]{EK3} for
basics of subfactor theory.

We introduce a notion of a symmetric bi-unitary connection.
Let $V$ be a finite set.  This is regarded as a common set of vertices of
a family of finite oriented graphs.
Let $G$ be a finite oriented graph whose vertices
are in $V$ and $\Delta_G$ its adjacency matrix.  That is,
$(\Delta_G)_{vw}$ is equal to the number of edges of $G$ from $v$ to $w$.
We assume that the matrix
$\Delta_G$ is symmetric and for a sufficiently large $n$,
all the entries of $\Delta_G^n$ are strictly positive.
Let $H$ be another finite oriented graph whose vertices are
again in $V$ and $\Delta_H$ be its adjacency matrix.  We assume
that for any $v\in V$, we have $w\in V$ with $(\Delta_H)_{vw}>0$
and that for any $w\in V$, we have $v\in V$ with $(\Delta_H)_{vw}>0$.
We assume that each vertex $v\in V$ has a positive value $\mu(v)$.  
We further assume to 
have two positive numbers $\be_G,\be_H$ such that for each $v$,
we have 
\begin{align*}
\be_G \mu(v)&= \sum_{w\in V} (\Delta_G)_{vw}\mu(w),\\
\be_H \mu(v)&= \sum_{w\in V} (\Delta_H)_{vw}\mu(w),\\
\be_H \mu(v)&= \sum_{w\in V} (\Delta_H)_{wv}\mu(w).
\end{align*}
That is, the vectors given by $\mu$ are the Perron-Frobenius
eigenvectors for $\Delta_G$, $\Delta_H$ and the transpose 
of $\Delta_H$.

We consider a \textit{cell} as in Fig. \ref{cell}, where the
edges $i,l$ are in $G$, the edges $j,k$ are in $H$, and we 
have $s(i)=s(k)$, $r(i)=s(j)$, $r(k)=s(l)$ and $r(j)=r(l)$.
(Here $s$ and $r$ denote the source and range of an oriented edge,
respectively.)

\thinlines
\unitlength 0.6mm
\begin{figure}[tb]
\begin{center}
\begin{picture}(55,30)
\multiput(11,10)(0,10){2}{\vector(1,0){8}}
\multiput(10,19)(10,0){2}{\vector(0,-1){8}}
\multiput(10,10)(10,0){2}{\circle*{1}}
\multiput(10,20)(10,0){2}{\circle*{1}}
\put(15,5){\makebox(0,0){$l$}}
\put(15,25){\makebox(0,0){$i$}}
\put(5,15){\makebox(0,0){$k$}}
\put(25,15){\makebox(0,0){$j$}}
\end{picture}
\end{center}
\caption{A cell}
\label{cell}
\end{figure}

We consider an assignment map $a$ of a complex number to
each cell and denote its value as in Fig. \ref{cell2}.
We sometimes draw a picture of a cell where at least one
of the requirements 
$s(i)=s(k)$, $r(i)=s(j)$, $r(k)=s(l)$ and $r(j)=r(l)$ fail.
In such a case, the picture in Fig. \ref{cell2} represents
a zero value.

\thinlines
\unitlength 0.6mm
\begin{figure}[tb]
\begin{center}
\begin{picture}(55,30)
\multiput(11,10)(0,10){2}{\vector(1,0){8}}
\multiput(10,19)(10,0){2}{\vector(0,-1){8}}
\multiput(10,10)(10,0){2}{\circle*{1}}
\multiput(10,20)(10,0){2}{\circle*{1}}
\put(15,5){\makebox(0,0){$l$}}
\put(15,25){\makebox(0,0){$i$}}
\put(5,15){\makebox(0,0){$k$}}
\put(25,15){\makebox(0,0){$j$}}
\put(15,15){\makebox(0,0){$a$}}
\end{picture}
\end{center}
\caption{A complex number}
\label{cell2}
\end{figure}
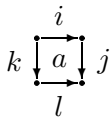

Consider a matrix $a$ defined as in Fig. \ref{matrix1},
where each row is labeled with a pair $(i,j)$ with $r(i)=s(j)$
and each column is labeled with a pair $(k,l)$ with $r(k)=s(l)$.
We use the same symbol $a$ for this matrix as the assignment map.
(Note that if we have $s(i)\neq s(k)$ or $r(j)\neq r(l)$, then
the entry is 0 by our convention.)
Then our first requirement is that this matrix $a$ is unitary.

\thinlines
\unitlength 0.6mm
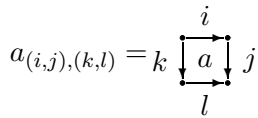
\begin{figure}[tb]
\begin{center}
\begin{picture}(55,30)
\multiput(36,10)(0,10){2}{\vector(1,0){8}}
\multiput(35,19)(10,0){2}{\vector(0,-1){8}}
\multiput(35,10)(10,0){2}{\circle*{1}}
\multiput(35,20)(10,0){2}{\circle*{1}}
\put(40,5){\makebox(0,0){$l$}}
\put(40,25){\makebox(0,0){$i$}}
\put(30,15){\makebox(0,0){$k$}}
\put(50,15){\makebox(0,0){$j$}}
\put(40,15){\makebox(0,0){$a$}}
\put(12,15){\makebox(0,0){$a_{(i,j),(k,l)}=$}}
\end{picture}
\end{center}
\caption{The first matrix}
\label{matrix1}
\end{figure}

Next consider a matrix $\bar a$ defined as in Fig. \ref{matrix2}, where
$\tilde j$ and $\tilde k$ denote the edges $j$ and $k$ with
reversed orientations and the bar above the square on the right
hand side means the complex conjugate.
Note that each row is labeled with a pair $(l,\tilde j)$ with $r(j)=r(l)$
and each column is labeled with a pair $(\tilde k,i)$ with $s(i)=s(k)$.
Then our second requirement is that the
matrix $\bar a$ is also unitary.  The two unitarity requirements
together are called \textsl{bi-unitarity}.  

\thinlines
\unitlength 0.6mm
\begin{figure}[tb]
\begin{center}
\begin{picture}(80,30)
\multiput(61,10)(0,10){2}{\vector(1,0){8}}
\multiput(60,19)(10,0){2}{\vector(0,-1){8}}
\multiput(60,10)(10,0){2}{\circle*{1}}
\multiput(60,20)(10,0){2}{\circle*{1}}
\put(65,5){\makebox(0,0){$l$}}
\put(65,25){\makebox(0,0){$i$}}
\put(55,15){\makebox(0,0){$k$}}
\put(75,15){\makebox(0,0){$j$}}
\put(65,15){\makebox(0,0){$a$}}
\put(12,15){\makebox(0,0){$\bar a_{(l,\tilde j),(\tilde k,i)}
\displaystyle=\sqrt{\frac{\mu(s(i))\mu(r(j))}{\mu(r(i))\mu(r(k))}}$}}
\put(56,30){\line(1,0){18}}
\end{picture}
\end{center}
\caption{The second matrix}
\label{matrix2}
\end{figure}

We further assume that we have the identity as in Fig. \ref{symmetric}.
This is called the \textsl{symmetric property} of $a$.

\thinlines
\unitlength 0.6mm
\begin{figure}[tb]
\begin{center}
\begin{picture}(100,30)
\multiput(11,10)(0,10){2}{\vector(1,0){8}}
\multiput(10,19)(10,0){2}{\vector(0,-1){8}}
\multiput(10,10)(10,0){2}{\circle*{1}}
\multiput(10,20)(10,0){2}{\circle*{1}}
\put(15,5){\makebox(0,0){$\tilde l$}}
\put(15,25){\makebox(0,0){$\tilde i$}}
\put(5,15){\makebox(0,0){$j$}}
\put(25,15){\makebox(0,0){$k$}}
\put(15,15){\makebox(0,0){$a$}}
\multiput(91,10)(0,10){2}{\vector(1,0){8}}
\multiput(90,19)(10,0){2}{\vector(0,-1){8}}
\multiput(90,10)(10,0){2}{\circle*{1}}
\multiput(90,20)(10,0){2}{\circle*{1}}
\put(95,5){\makebox(0,0){$l$}}
\put(95,25){\makebox(0,0){$i$}}
\put(85,15){\makebox(0,0){$k$}}
\put(105,15){\makebox(0,0){$j$}}
\put(95,15){\makebox(0,0){$a$}}
\put(56,15){\makebox(0,0)
{$\displaystyle=\sqrt{\frac{\mu(s(i))\mu(r(j))}{\mu(r(i))\mu(r(k))}}$}}
\put(86,30){\line(1,0){18}}
\end{picture}
\end{center}
\caption{The symmetric property of $a$}
\label{symmetric}
\end{figure}

If all of the above are satisfied, we say that $a$ is a \textsl{symmetric
bi-unitary connection} on $G,H$.

Let $a$ be a symmetric bi-unitary connection on $G,H$ as above.
Let $U=(U_{j'j})$ be a unitary matrix where $j,j'$
are edges of the graph $H$.  We assume that $U_{j'j}=0$
if $s(j)\neq s(j')$ or $r(j)\neq r(j')$.
For such $a,U$, we set a new symmetric bi-unitary connection $\tilde a$
as in Fig. \ref{equiv}.  In this case, we say the symmetric
bi-unitary connections $\tilde a$ and $a$ are \textsl{equivalent}.

\thinlines
\unitlength 0.6mm
\begin{figure}[tb]
\begin{center}
\begin{picture}(105,30)
\multiput(11,10)(0,10){2}{\vector(1,0){8}}
\multiput(10,19)(10,0){2}{\vector(0,-1){8}}
\multiput(10,10)(10,0){2}{\circle*{1}}
\multiput(10,20)(10,0){2}{\circle*{1}}
\put(15,5){\makebox(0,0){$l$}}
\put(15,25){\makebox(0,0){$i$}}
\put(5,15){\makebox(0,0){$k$}}
\put(25,15){\makebox(0,0){$j$}}
\put(15,15){\makebox(0,0){$\tilde a$}}
\multiput(71,10)(0,10){2}{\vector(1,0){8}}
\multiput(70,19)(10,0){2}{\vector(0,-1){8}}
\multiput(70,10)(10,0){2}{\circle*{1}}
\multiput(70,20)(10,0){2}{\circle*{1}}
\put(75,5){\makebox(0,0){$l$}}
\put(75,25){\makebox(0,0){$i$}}
\put(65,15){\makebox(0,0){$k'$}}
\put(85,15){\makebox(0,0){$j'$}}
\put(75,15){\makebox(0,0){$a$}}
\put(45,12){\makebox(0,0){$=\displaystyle\sum_{j',k'}\bar U_{k'k}$}}
\put(95,15){\makebox(0,0){$U_{j'j}$}}
\end{picture}
\end{center}
\caption{Equivalence of symmetric bi-unitary connections}
\label{equiv}
\end{figure}

Let $a$ be a symmetric bi-unitary connection on $G,H_1$ and
$b$ be a symmetric bi-unitary connection on $G,H_2$.
We define a new graph $H_3$ simply by adding the edges of
$H_1$ and $H_2$.  That is, the adjacency matrix is given
by $\Delta_{H_3}=\Delta_{H_1}+\Delta_{H_2}$.
We can define a new symmetric bi-unitary connection $c$
by setting a matrix $c_{(i,j),(k,l)}$ as follows.
\[
c_{(i,j),(k,l)}=
\begin{cases}
a_{(i,j),(k,l)},&\text{if } j,k\in H_1,\\
b_{(i,j),(k,l)},&\text{if } j,k\in H_2,\\
0, &\text{otherwise.}
\end{cases}
\]
Note that we have $\be_{H_3}=\be_{H_1}+\be_{H_2}$.
We call $c$ a sum of $a$ and $b$, and write $c=a+b$.

We say a symmetric bi-unitary connection is
\textsl{irreducible} if it is not equivalent to a sum
of two symmetric bi-unitary connections.

Let $a$ be a symmetric bi-unitary connection on $G, H_1$ and $b$ 
be a symmetric bi-unitary connection on $G, H_2$. We define a new 
graph $H_3$ simply by concatenating the graphs $H_1$ and $H_2$
vertically. That is, the adjacency matrix is given by 
$\Delta_{H_3} = \Delta_{H_1}\Delta_{H_2}$.
We define a new symmetric bi-unitary connection $ab$ on
$G, H_3$ as in Figures \ref{prod1} and \ref{prod2}.

\thinlines
\unitlength 0.6mm
\begin{figure}[tb]
\begin{center}
\begin{picture}(105,40)
\multiput(11,10)(0,20){2}{\vector(1,0){8}}
\multiput(10,19)(10,0){2}{\vector(0,-1){8}}
\multiput(10,29)(10,0){2}{\vector(0,-1){8}}
\multiput(10,10)(10,0){2}{\circle*{1}}
\multiput(10,20)(10,0){2}{\circle*{1}}
\multiput(10,30)(10,0){2}{\circle*{1}}
\put(15,5){\makebox(0,0){$n$}}
\put(15,35){\makebox(0,0){$i$}}
\put(5,15){\makebox(0,0){$m$}}
\put(5,25){\makebox(0,0){$l$}}
\put(25,15){\makebox(0,0){$k$}}
\put(25,25){\makebox(0,0){$j$}}
\put(15,25){\makebox(0,0){$a$}}
\put(15,15){\makebox(0,0){$b$}}
\multiput(56,15)(0,10){2}{\vector(1,0){8}}
\multiput(55,24)(10,0){2}{\vector(0,-1){8}}
\multiput(55,15)(10,0){2}{\circle*{1}}
\multiput(55,25)(10,0){2}{\circle*{1}}
\put(60,10){\makebox(0,0){$o$}}
\put(60,30){\makebox(0,0){$i$}}
\put(50,20){\makebox(0,0){$l$}}
\put(70,20){\makebox(0,0){$j$}}
\put(60,20){\makebox(0,0){$a$}}
\multiput(81,15)(0,10){2}{\vector(1,0){8}}
\multiput(80,24)(10,0){2}{\vector(0,-1){8}}
\multiput(80,15)(10,0){2}{\circle*{1}}
\multiput(80,25)(10,0){2}{\circle*{1}}
\put(85,10){\makebox(0,0){$n$}}
\put(85,30){\makebox(0,0){$o$}}
\put(75,20){\makebox(0,0){$m$}}
\put(95,20){\makebox(0,0){$k$}}
\put(85,20){\makebox(0,0){$b$}}
\put(38,18){\makebox(0,0){$\displaystyle=\sum_o$}}
\end{picture}
\end{center}
\caption{The product of connections $a,b$}
\label{prod1}
\end{figure}

\thinlines
\unitlength 0.6mm
\begin{figure}[tb]
\begin{center}
\begin{picture}(75,40)
\multiput(16,15)(0,10){2}{\vector(1,0){8}}
\multiput(15,24)(10,0){2}{\vector(0,-1){8}}
\multiput(15,15)(10,0){2}{\circle*{1}}
\multiput(15,25)(10,0){2}{\circle*{1}}
\put(20,10){\makebox(0,0){$n$}}
\put(20,30){\makebox(0,0){$i$}}
\put(5,20){\makebox(0,0){$(l,m)$}}
\put(35,20){\makebox(0,0){$(j,k)$}}
\put(20,20){\makebox(0,0){$ab$}}
\multiput(56,10)(0,20){2}{\vector(1,0){8}}
\multiput(56,19)(10,0){2}{\vector(0,-1){8}}
\multiput(56,29)(10,0){2}{\vector(0,-1){8}}
\multiput(56,10)(10,0){2}{\circle*{1}}
\multiput(56,20)(10,0){2}{\circle*{1}}
\multiput(56,30)(10,0){2}{\circle*{1}}
\put(60,5){\makebox(0,0){$n$}}
\put(60,35){\makebox(0,0){$i$}}
\put(50,15){\makebox(0,0){$m$}}
\put(50,25){\makebox(0,0){$l$}}
\put(70,15){\makebox(0,0){$k$}}
\put(70,25){\makebox(0,0){$j$}}
\put(60,25){\makebox(0,0){$a$}}
\put(60,15){\makebox(0,0){$b$}}
\put(45,20){\makebox(0,0){$=$}}
\end{picture}
\end{center}
\caption{The product of connections $a,b$}
\label{prod2}
\end{figure}

For a symmetric bi-unitary connection $a$, we also 
define a new one $\bar a$ as in Fig. \ref{abar}.
This is called the \textsl{dual} 
symmetric bi-unitary connection of $a$.

\thinlines
\unitlength 0.6mm
\begin{figure}[tb]
\begin{center}
\begin{picture}(100,30)
\multiput(11,10)(0,10){2}{\vector(1,0){8}}
\multiput(10,19)(10,0){2}{\vector(0,-1){8}}
\multiput(10,10)(10,0){2}{\circle*{1}}
\multiput(10,20)(10,0){2}{\circle*{1}}
\put(15,5){\makebox(0,0){$i$}}
\put(15,25){\makebox(0,0){$l$}}
\put(5,15){\makebox(0,0){$\tilde k$}}
\put(25,15){\makebox(0,0){$\tilde j$}}
\put(15,15){\makebox(0,0){$\bar a$}}
\multiput(91,10)(0,10){2}{\vector(1,0){8}}
\multiput(90,19)(10,0){2}{\vector(0,-1){8}}
\multiput(90,10)(10,0){2}{\circle*{1}}
\multiput(90,20)(10,0){2}{\circle*{1}}
\put(95,5){\makebox(0,0){$l$}}
\put(95,25){\makebox(0,0){$i$}}
\put(85,15){\makebox(0,0){$k$}}
\put(105,15){\makebox(0,0){$j$}}
\put(95,15){\makebox(0,0){$a$}}
\put(56,15){\makebox(0,0)
{$\displaystyle=\sqrt{\frac{\mu(s(i))\mu(r(j))}{\mu(r(i))\mu(r(k))}}$}}
\put(86,30){\line(1,0){18}}
\end{picture}
\end{center}
\caption{The new connection $\bar a$}
\label{abar}
\end{figure}

At the end of this section, we recall relations between symmetric
bi-unitary connections and bimodules over II$_1$ factors.
(See \cite[Chapter 9]{EK3} for general theory of bimodules.)
We fix a vertex $*$ in $V$.  Using the graph $G$, we obtain
a hyperfinite II$_1$ factor $M$ with the string algebra
construction as in \cite[Section 11.3]{EK3}.
From a symmetric bi-unitary connection $a$, we obtain
an $M$-$M$ bimodule $H_a$ with the open string bimodule
construction as in \cite{S}, \cite[Section 3]{AH}.  Note that here
we have $\dim_M H_a<\infty$ and $\dim (H_a)_M<\infty$.
We recall the following result in \cite[Section 3]{AH}.
(Note that here we use only \textsl{symmetric} bi-unitary 
connections here.)

\begin{theorem}\label{conn-bim}
For symmetric bi-unitary connections $a,b$,
we have $H_{a+b}\isom H_a\oplus H_b$,
$H_{ab}\isom H_a\otimes_M H_b$ and
$\bar H_a\isom H_{\bar a}$.  These are $M$-$M$
bimodule isomorphisms.  

We have $H_a\isom H_b$ if and only if $a$ and $b$ are
equivalent.  We also have $a$ is
irreducible if and only if $H_a$ is an irreducible
$M$-$M$ bimodule.
\end{theorem}

Let $a$ be a symmetric bi-unitary connection on $G,H_1$ and
$b$ be a symmetric bi-unitary connection on $G,H_2$.
By the results in \cite[Section 3]{AH}, we can describe
the bimodule homomorphism space $\Hom(H_a, H_b)$ in terms
of symmetric bi-unitary connections.  

\section{Tensors, flat connections and matrix product operator algebras}

We next introduce a family of flat bi-unitary connections arising
from a subfactor.  Let $N\subset M$ be a subfactor
(of the hyperfinite II$_1$ factor with finite Jones index and
finite depth) as in \cite[Chapter 9]{EK3}.
We do \textsl{not} assume the irreducibility $N'\cap M=\CC$ here.
We set  $N=M_{-1}$, $M=M_0$ and apply the
Jones tower/tunnel construction \cite[Section 9.3]{EK3} to get
$$\cdots\subset M_{-2}\subset M_{-1}\subset M_0\subset M_1
\subset M_2\subset\cdots.$$  Each algebra $M_j$ is again a
type II$_1$ factor.  We then consider a double sequence 
$A_{jk}=M'_{-k}\cap M_j$ of finite dimensional $C^*$-algebras
for $j,k\ge0$.  They form \textsl{commuting squares} as in
\cite[Chapter 9]{EK3}.  We label minimal direct summands
of $M'_{-1}\cap M_{2j+1}$ with labels $a,b,c,\dots$.
The labels of those for $M'_{-1}\cap M_{2j+1}$ 
are naturally identified with those for $M'_{-1}\cap M_{2j+3}$.
Fix a label $a$ and choose a minimal projection $p$ from the
summand of $M'_{-1}\cap M_{2j+1}$ 
labeled with $a$ and consider the sequence of commuting square
\[
\begin{array}{cccccccc}
(M'_{-1}\cap M_{-1})p & \subset & (M'_{-3}\cap M_{-1})p & \subset &
(M'_{-5}\cap M_{-1})p & \subset & (M'_{-7}\cap M_{-1})p &\subset\cdots \\
\cap && \cap && \cap && \cap &\\
p(M'_{-1}\cap M_{2j+1})p & \subset & p(M'_{-3}\cap M_{2j+1})p & \subset &
p(M'_{-5}\cap M_{2j+1})p & \subset & p(M'_{-7}\cap M_{2j+1})p
&\subset\cdots. 
\end{array}
\]

\begin{figure}[tb]
\thinlines
\unitlength 0.6mm
\begin{center}
\begin{picture}(202,75)
\put(11,10){\vector(1,0){8}}
\put(21,10){\vector(1,0){8}}
\put(51,10){\vector(1,0){8}}
\put(11,60){\vector(1,0){8}}
\put(21,60){\vector(1,0){8}}
\put(51,60){\vector(1,0){8}}
\put(10,19){\vector(0,-1){8}}
\put(10,49){\vector(0,-1){8}}
\put(10,59){\vector(0,-1){8}}
\put(60,19){\vector(0,-1){8}}
\put(60,49){\vector(0,-1){8}}
\put(60,59){\vector(0,-1){8}}
\multiput(10,39)(0,-4){5}{\line(0,-1){3}}
\multiput(60,39)(0,-4){5}{\line(0,-1){3}}
\multiput(31,10)(4,0){5}{\line(1,0){3}}
\multiput(31,60)(4,0){5}{\line(1,0){3}}
\put(20,10){\circle*{1}}
\put(30,10){\circle*{1}}
\put(50,10){\circle*{1}}
\put(20,60){\circle*{1}}
\put(30,60){\circle*{1}}
\put(50,60){\circle*{1}}
\put(10,20){\circle*{1}}
\put(10,40){\circle*{1}}
\put(10,50){\circle*{1}}
\put(60,20){\circle*{1}}
\put(60,40){\circle*{1}}
\put(60,50){\circle*{1}}
\put(10,10){\makebox(0,0){$*$}}
\put(10,60){\makebox(0,0){$*$}}
\put(60,10){\makebox(0,0){$*$}}
\put(60,60){\makebox(0,0){$*$}}
\put(15,65){\makebox(0,0){$k_1$}}
\put(25,65){\makebox(0,0){$k_2$}}
\put(55,65){\makebox(0,0){$k_n$}}
\put(15,5){\makebox(0,0){$j_1$}}
\put(25,5){\makebox(0,0){$j_2$}}
\put(55,5){\makebox(0,0){$j_{n}$}}
\put(3,15){\makebox(0,0){$i_{2m}$}}
\put(5,45){\makebox(0,0){$i_{2}$}}
\put(5,55){\makebox(0,0){$i_{1}$}}
\put(67,15){\makebox(0,0){$l_{2m}$}}
\put(65,45){\makebox(0,0){$l_{2}$}}
\put(65,55){\makebox(0,0){$l_{1}$}}
\put(15,15){\makebox(0,0){$\bar a$}}
\put(25,15){\makebox(0,0){$\bar a$}}
\put(55,15){\makebox(0,0){$\bar a$}}
\put(15,45){\makebox(0,0){$\bar a$}}
\put(25,45){\makebox(0,0){$\bar a$}}
\put(55,45){\makebox(0,0){$\bar a$}}
\put(15,55){\makebox(0,0){$a$}}
\put(25,55){\makebox(0,0){$a$}}
\put(55,55){\makebox(0,0){$a$}}
\put(72,33){$=\de_{j_1,k_1}
\de_{j_2,k_2}\cdots
\de_{j_{n},k_{n}}
\de_{i_1,l_1}
\de_{i_2,l_2}\cdots
\de_{i_{2m},{l_{2m}}}.$}
\end{picture}
\end{center}
\caption{Flatness property}
\label{flat}
\end{figure}
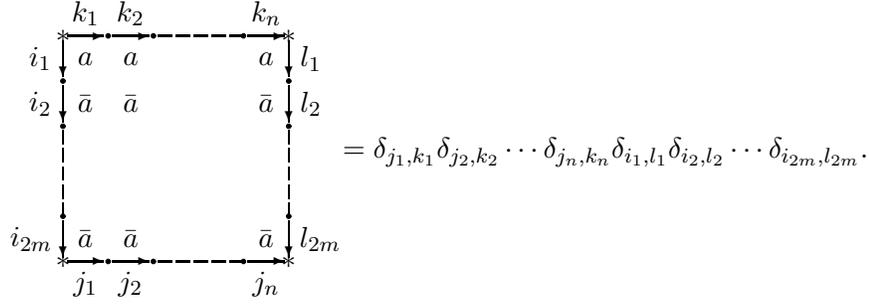

We denote the symmetric bi-unitary connection which gives these
commuting squares by $a$.  (Since we use only $M_{2j+1}$, this 
we have the symmetric property.) The symmetric bi-unitary 
connection $a$ is well-defined up to equivalence
regardless of the choices of $j$ and $p$.  Furthermore, this satisfies
\textsl{flatness} as in Fig. \ref{flat}.  
(See \cite{O1}, \cite{K1},
\cite[Chapter 10]{EK3}. The name ``flatness'' comes from
analogy to a flat connection in differential geometry.
A graph is regarded as a discrete analogue of a manifold.)

Since we have the finite depth assumption, we have
finitely many labels $a,b,c,\dots$ for irreducible
flat symmetric
bi-unitary connections arising from $N\subset M$.
By Theorem \ref{conn-bim} and the remark after that,
we have a fusion category whose objects are labeled
with $a,b,c,\dots$ and this fusion category is
equivalent to that of $N$-$N$ bimodules arising from
the subfactor $N\subset M$.  (See \cite{K2} for
general theory of fusion categories and its relation
to operator algebras.)  Each vertex $v$ in $V$ corresponds
to an irreducible $N$-$N$ bimodule $H_v$ arising from 
the subfactor $N\subset M$ and $\mu(v)$ is given by the
(quantum) dimension of $H_v$.
One advantage of using flat symmetric
bi-unitary connections is that all information can be
encoded into a single matrix (by making a direct sum of
matrices) while we have to with infinite dimensional
Hilbert spaces in the bimodule approach to fusion categories
(or endomorphisms of an infinite dimensional von Neumann
algebras in the sector approach \cite{L}).

Note that the flat symmetric
bi-unitary connection labeled with $a$ is given
by the quantum $6j$-symbols for the $N$-$N$ bimodules
arising from the subfactor $N\subset M$
where two of the six bimodules are labeled with $a$ 
and given by ${}_N M_N$.  (See \cite[Chapter 12]{EK3}
for quantum $6j$-symbols arising from subfactors.
The flatness condition roughly corresponds to the
pentagon relations of quantum $6j$-symbols as
explained in \cite[Chapter 12]{EK3}.)

We now relate the above definitions to the setting of
\cite{BMWSHV}.
We define a tensor $a$ with four legs as in Fig. \ref{tensor}.
(We use the same label for a tensor and a flat symmetric bi-unitary
connection.)  Recall that we keep the convention that if 
$s(i)\neq s(k)$, 
$r(i)\neq s(j)$, $r(k)\neq s(l)$ or $r(j)\neq r(l)$, we have
the value $0$.  Note that if the fusion category is realized with
a finite group with a 3-cocycle, then all the normalizing
constants in Fig. \ref{tensor} are $1$.

\thinlines
\unitlength 0.6mm
\begin{figure}[tb]
\begin{center}
\begin{picture}(120,30)
\put(15,15){\circle{6}}
\put(18,15){\line(1,0){3}}
\put(9,15){\line(1,0){3}}
\put(15,18){\line(0,1){3}}
\put(15,9){\line(0,1){3}}
\multiput(96,10)(0,10){2}{\vector(1,0){8}}
\multiput(95,19)(10,0){2}{\vector(0,-1){8}}
\multiput(95,10)(10,0){2}{\circle*{1}}
\multiput(95,20)(10,0){2}{\circle*{1}}
\put(15,15){\makebox(0,0){$a$}}
\put(25,15){\makebox(0,0){$j$}}
\put(5,15){\makebox(0,0){$k$}}
\put(15,25){\makebox(0,0){$i$}}
\put(15,5){\makebox(0,0){$l$}}
\put(100,5){\makebox(0,0){$l$}}
\put(100,25){\makebox(0,0){$i$}}
\put(90,15){\makebox(0,0){$k$}}
\put(110,15){\makebox(0,0){$j$}}
\put(100,15){\makebox(0,0){$a$}}
\put(58,15){\makebox(0,0){$\displaystyle=\sqrt[4]
{\frac{\mu(s(i))\mu(r(j))}{\mu(r(i))\mu(r(k))}}$}}
\end{picture}
\end{center}
\caption{A tensor defined from a symmetric bi-unitary connection}
\label{tensor}
\end{figure}

It is easy to see that the tensor $ab$ 
corresponding to the product of the two flat symmetric
bi-unitary connections $a$ and $b$ is
given by the vertical concatenation 
and contraction of the tensors $a$ and $b$, and 
we have Fig. \ref{tensorbar} for the tensor $\bar a$
corresponding the dual flat symmetric bi-unitary connection.
(Note that the latter holds due to our convention of
the normalizing constants as in Fig. \ref{tensor}.)

\thinlines
\unitlength 0.6mm
\begin{figure}[tb]
\begin{center}
\begin{picture}(70,30)
\put(15,15){\circle{6}}
\put(18,15){\line(1,0){3}}
\put(9,15){\line(1,0){3}}
\put(15,18){\line(0,1){3}}
\put(15,9){\line(0,1){3}}
\put(15,15){\makebox(0,0){$\bar a$}}
\put(25,15){\makebox(0,0){$j$}}
\put(5,15){\makebox(0,0){$k$}}
\put(15,25){\makebox(0,0){$i$}}
\put(15,5){\makebox(0,0){$l$}}
\put(30,15){\makebox(0,0){$=$}}
\put(45,15){\circle{6}}
\put(48,15){\line(1,0){3}}
\put(39,15){\line(1,0){3}}
\put(45,18){\line(0,1){3}}
\put(45,9){\line(0,1){3}}
\put(45,15){\makebox(0,0){$a$}}
\put(55,15){\makebox(0,0){$\tilde j$}}
\put(35,15){\makebox(0,0){$\tilde k$}}
\put(45,25){\makebox(0,0){$l$}}
\put(45,5){\makebox(0,0){$i$}}
\put(33,30){\line(1,0){24}}
\end{picture}
\end{center}
\caption{A tensor $\bar a$}
\label{tensorbar}
\end{figure}

Then the zipper condition, which is assumed  in 
\cite[(2)]{BMWSHV}, now holds for our setting as in Fig. \ref{zipper}.
This is due to the irreducible decomposition of the product
of two flat symmetric bi-unitary connections.

The advantage of using {\sl{flat}} symmetric bi-unitary connections
is that we get (a part of) the irreducible decomposition rules
canonically from the (dual) principal graphs.  The zipper
condition holds for general symmetric bi-unitary connections
even without flatness, but then it is very difficult to see
the irreducible decomposition rules and there is a possibility
that we need infinitely many tensors and do not have a fusion
category, which happens when the subfactor has an infinite
depth.

\thinlines
\unitlength 0.6mm
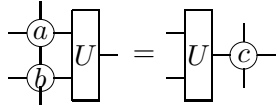
\begin{figure}[tb]
\begin{center}
\begin{picture}(70,30)
\put(15,15){\circle{6}}
\put(15,25){\circle{6}}
\put(18,15){\line(1,0){4}}
\put(18,25){\line(1,0){4}}
\put(8,15){\line(1,0){4}}
\put(8,25){\line(1,0){4}}
\put(28,20){\line(1,0){4}}
\put(15,28){\line(0,1){4}}
\put(15,18){\line(0,1){4}}
\put(15,9){\line(0,1){4}}
\put(22,10){\line(1,0){6}}
\put(22,30){\line(1,0){6}}
\put(22,10){\line(0,1){20}}
\put(28,10){\line(0,1){20}}
\put(43,15){\line(1,0){4}}
\put(43,25){\line(1,0){4}}
\put(53,20){\line(1,0){4}}
\put(63,20){\line(1,0){4}}
\put(47,10){\line(1,0){6}}
\put(47,30){\line(1,0){6}}
\put(47,10){\line(0,1){20}}
\put(53,10){\line(0,1){20}}
\put(60,20){\circle{6}}
\put(60,13){\line(0,1){4}}
\put(60,23){\line(0,1){4}}
\put(15,15){\makebox(0,0){$b$}}
\put(15,25){\makebox(0,0){$a$}}
\put(60,20){\makebox(0,0){$c$}}
\put(25,20){\makebox(0,0){$U$}}
\put(50,20){\makebox(0,0){$U$}}
\put(38,20){\makebox(0,0){$=$}}
\end{picture}
\end{center}
\caption{The zipper condition}
\label{zipper}
\end{figure}

Next we introduce a \textsl{matrix product state}.
We concatenate $L$ tensors with 3 legs and use the
value of the contraction for a state labeled with
$i_1,i_2,\dots,i_L$ where these indices label
tensors.  See Fig. \ref{mps} where $L=4$.
A matrix product state was first introduced in
\cite{FNW} and has been important in recent studies
of gapped Hamiltonians.

\thinlines
\unitlength 0.6mm
\begin{figure}[tb]
\begin{center}
\begin{picture}(120,40)
\put(40,25){\circle{6}}
\put(50,25){\circle{6}}
\put(60,25){\circle{6}}
\put(70,25){\circle{6}}
\put(33,25){\line(1,0){4}}
\put(43,25){\line(1,0){4}}
\put(53,25){\line(1,0){4}}
\put(63,25){\line(1,0){4}}
\put(73,25){\line(1,0){4}}
\put(33,10){\line(1,0){44}}
\put(33,17.5){\arc{15}{1.571}{4.713}}
\put(77,17.5){\arc{15}{4.713}{7.855}}
\put(40,28){\line(0,1){4}}
\put(50,28){\line(0,1){4}}
\put(60,28){\line(0,1){4}}
\put(70,28){\line(0,1){4}}
\put(40,36){\makebox(0,0){$i_1$}}
\put(50,36){\makebox(0,0){$i_2$}}
\put(60,36){\makebox(0,0){$i_3$}}
\put(70,36){\makebox(0,0){$i_4$}}
\put(40,25){\makebox(0,0){$a$}}
\put(50,25){\makebox(0,0){$a$}}
\put(60,25){\makebox(0,0){$a$}}
\put(70,25){\makebox(0,0){$a$}}
\put(100,18){\makebox(0,0){$|i_1i_2i_3i_4\rangle$}}
\put(10,18){\makebox(0,0){$\displaystyle\sum_{i_1,i_2,i_3,i_4}$}}
\end{picture}
\end{center}
\caption{A matrix product state}
\label{mps}
\end{figure}
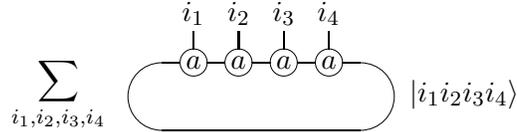

In a similar way, we introduce a \textsl{matrix product
operator} by using $L$ tensors with 4 legs as in 
Fig. \ref{mpo}, where $L$ is again 4.  When we use the
tensor labeled with $a$ arising from a flat
symmetric bi-unitary connection, we
denote the resulting matrix product operator by
$O^L_a$.

\thinlines
\unitlength 0.6mm
\begin{figure}[tb]
\begin{center}
\begin{picture}(160,40)
\put(50,25){\circle{6}}
\put(60,25){\circle{6}}
\put(70,25){\circle{6}}
\put(80,25){\circle{6}}
\put(43,25){\line(1,0){4}}
\put(53,25){\line(1,0){4}}
\put(63,25){\line(1,0){4}}
\put(73,25){\line(1,0){4}}
\put(83,25){\line(1,0){4}}
\put(43,10){\line(1,0){44}}
\put(43,17.5){\arc{15}{1.571}{4.713}}
\put(87,17.5){\arc{15}{4.713}{7.855}}
\put(50,18){\line(0,1){4}}
\put(60,18){\line(0,1){4}}
\put(70,18){\line(0,1){4}}
\put(80,18){\line(0,1){4}}
\put(50,28){\line(0,1){4}}
\put(60,28){\line(0,1){4}}
\put(70,28){\line(0,1){4}}
\put(80,28){\line(0,1){4}}
\put(50,14){\makebox(0,0){$j_1$}}
\put(60,14){\makebox(0,0){$j_2$}}
\put(70,14){\makebox(0,0){$j_3$}}
\put(80,14){\makebox(0,0){$j_4$}}
\put(50,36){\makebox(0,0){$i_1$}}
\put(60,36){\makebox(0,0){$i_2$}}
\put(70,36){\makebox(0,0){$i_3$}}
\put(80,36){\makebox(0,0){$i_4$}}
\put(50,25){\makebox(0,0){$a$}}
\put(60,25){\makebox(0,0){$a$}}
\put(70,25){\makebox(0,0){$a$}}
\put(80,25){\makebox(0,0){$a$}}
\put(125,18){\makebox(0,0){$|i_1i_2i_3i_4\rangle\langle j_1j_2j_3j_4|$}}
\put(10,18){\makebox(0,0){$\displaystyle
\sum_{i_1,i_2,i_3,i_4,j_1,j_2,j_3,j_4}$}}
\end{picture}
\end{center}
\caption{A matrix product operator}
\label{mpo}
\end{figure}
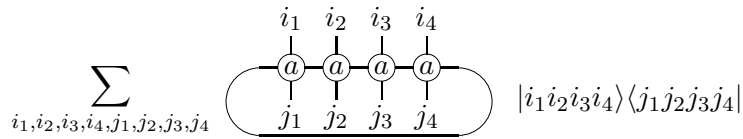

The product of two matrix product operators $O^L_a$, $O^L_b$ is
given as in Fig. \ref{pmpo}.

\thinlines
\unitlength 0.6mm
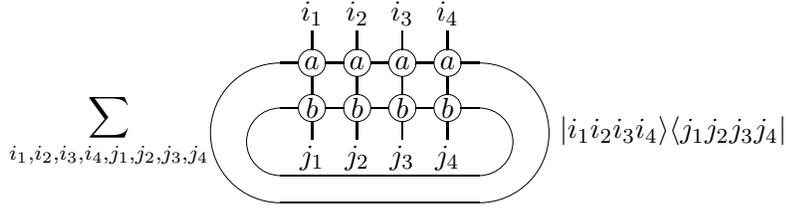
\begin{figure}[tb]
\begin{center}
\begin{picture}(165,60)
\put(50,35){\circle{6}}
\put(60,35){\circle{6}}
\put(70,35){\circle{6}}
\put(80,35){\circle{6}}
\put(50,45){\circle{6}}
\put(60,45){\circle{6}}
\put(70,45){\circle{6}}
\put(80,45){\circle{6}}
\put(43,35){\line(1,0){4}}
\put(53,35){\line(1,0){4}}
\put(63,35){\line(1,0){4}}
\put(73,35){\line(1,0){4}}
\put(83,35){\line(1,0){4}}
\put(43,45){\line(1,0){4}}
\put(53,45){\line(1,0){4}}
\put(63,45){\line(1,0){4}}
\put(73,45){\line(1,0){4}}
\put(83,45){\line(1,0){4}}
\put(43,20){\line(1,0){44}}
\put(43,14){\line(1,0){44}}
\put(43,27.5){\arc{15}{1.571}{4.713}}
\put(87,27.5){\arc{15}{4.713}{7.855}}
\put(43,29.5){\arc{31}{1.571}{4.713}}
\put(87,29.5){\arc{31}{4.713}{7.855}}
\put(50,28){\line(0,1){4}}
\put(60,28){\line(0,1){4}}
\put(70,28){\line(0,1){4}}
\put(80,28){\line(0,1){4}}
\put(50,38){\line(0,1){4}}
\put(60,38){\line(0,1){4}}
\put(70,38){\line(0,1){4}}
\put(80,38){\line(0,1){4}}
\put(50,48){\line(0,1){4}}
\put(60,48){\line(0,1){4}}
\put(70,48){\line(0,1){4}}
\put(80,48){\line(0,1){4}}
\put(50,24){\makebox(0,0){$j_1$}}
\put(60,24){\makebox(0,0){$j_2$}}
\put(70,24){\makebox(0,0){$j_3$}}
\put(80,24){\makebox(0,0){$j_4$}}
\put(50,56){\makebox(0,0){$i_1$}}
\put(60,56){\makebox(0,0){$i_2$}}
\put(70,56){\makebox(0,0){$i_3$}}
\put(80,56){\makebox(0,0){$i_4$}}
\put(50,35){\makebox(0,0){$b$}}
\put(60,35){\makebox(0,0){$b$}}
\put(70,35){\makebox(0,0){$b$}}
\put(80,35){\makebox(0,0){$b$}}
\put(50,45){\makebox(0,0){$a$}}
\put(60,45){\makebox(0,0){$a$}}
\put(70,45){\makebox(0,0){$a$}}
\put(80,45){\makebox(0,0){$a$}}
\put(130,30){\makebox(0,0){$|i_1i_2i_3i_4\rangle\langle j_1j_2j_3j_4|$}}
\put(5,30){\makebox(0,0){$\displaystyle
\sum_{i_1,i_2,i_3,i_4,j_1,j_2,j_3,j_4}$}}
\end{picture}
\end{center}
\caption{A product of matrix product operators}
\label{pmpo}
\end{figure}

By the zipper condition in Fig. \ref{zipper}, the
product $O^L_a O^L_b$ decomposes as $\sum_c N_{ab}^c O^L_c$,
where the coefficients $N_{ab}^c$'s are  given in the
decomposition of the product $ab$ of flat symmetric
bi-unitary connections into $\sum_c N_{ab}^c$.
We also see that $O^L_a$'s give a fusion category where
the dual object of $O^L_a$ is $O^L_{\bar a}$.  Note that
this fusion category is determined uniquely up to
equivalence regardless of $L$.  This is clear from the
zipper condition, Fig. \ref{zipper}.  From a subfactor
theory viewpoint, this is because horizontal basic
construction of a commuting square does not change
a resulting subfactor.  Also, we have $P_L=\sum_a w_a O^L$
as in \cite[(13)]{BMWSHV}, where $w_a=d_a/w$, $w=\sum_a d_a^2$,
and $d_a$ is the Perron-Frobenius dimension of $a$.

We next consider Ocneanu's tube algebra \cite{EK2}, 
\cite[Section 12.6]{EK3},
\cite{I}, \cite{KSW}.  We recall the definition following \cite{I}.

\begin{definition}{\rm
Let $\C$ be a unitary fusion category.
We set the tube algebra $\Tube(\C)$ to be
\[
\bigoplus_{\la,\nu\in\Irr(\C),\mu\in\Irr(\C)}
\Hom(\la\mu,\mu\nu)
\]
as a complex linear space.  We define its algebra structure and
$*$-structure by the formulas as in \cite[page 134]{I}.
}\end{definition}

We then see that under the identification of flat symmetric
bi-unitary connections and tensors as in Fig. \ref{tensor},
the tube algebra arising from the unitary fusion category
of the flat symmetric bi-unitary connections and the anyon
algebra defined for the tensors as in \cite[page 199]{BMWSHV}
are isomorphic.  We thus have the following result.

\begin{theorem}
A family of tensors arising from the
bi-unitary flat connections as above satisfy all
the requirements in 
Bultinck-Mari\"ena-Williamson-\c Sahino\u glu-Haegemana-Verstraete
\cite{BMWSHV}.  The resulting anyon algebra and
the tube algebra for the subfactor are isomorphic.
The tensor categories arising from these two algebras
are the equivalent, hence both are modular tensor categories
and the Verlinde formula holds for the former.
\end{theorem}

That is, the new tensor category arising from the anyon algebra
in \cite{BMWSHV} is indeed the Drinfel$'$d center of the
original unitary fusion category.
They define the fusion rules of the new tensor category in 
\cite{BMWSHV}, but they say ``The multiplicities --- the 
specific values of $N^k_{ij}$ --- are in general harder to 
obtain directly since they arise from the number of linearly 
independent ways the MPO strings emanating from the
idempotents can be connected on the virtual level.''
So it is not clear whether the structure constants $N_{ij}^k$
are well-defined or not in \cite{BMWSHV}, but these are
well-defined by \cite{EK3}, \cite{I}, \cite{KSW}.
In \cite{BMWSHV}, they also say ``One could of course also
just calculate the fusion
multiplicities from the $S$ matrix using the Verlinde formula,''
but it is not clear
why the Verlinde formula applies to this setting.  Now
our identification clearly shows that the Verlinde formula
applies as in \cite[Section 12.7]{EK3}, \cite{I}, \cite{KSW}.

\section{Non-flat connections and their flat parts}

We have seen a family of flat symmetric bi-unitary connections
arising from a subfactor produces tensors satisfying the
requirement of \cite{BMWSHV}, but more general tensors also satisfy 
that.  We now look into this matter.

Consider the Dynkin diagram $E_7$.  It has a bi-unitary connection
and this is not flat as in \cite{EK1}, \cite[Chapter 11]{EK3}.
Though this bi-unitary connection is not symmetric, it is easy
to obtain a symmetric one through horizontal and vertical
basic constructions.  Starting with this symmetric bi-unitary
connection, we get a family of irreducible symmetric bi-unitary
connections.  (The set $V$ consists of the four even vertices of
$E_7$.)  This situation was considered by Ocneanu \cite{O2}.
Since the principal graph of the subfactor arising from the
bi-unitary connection on $E_7$ is $D_{10}$ as in \cite{EK1},
we obtain 6 irreducible symmetric bi-unitary connections
in this way.  (The number 6 is give by the even vertices
of $D_{10}$.)  Actually, one can obtain a larger fusion category
of bi-unitary connections as in Ocneanu \cite{O2} for $E_7$.
He has 17 irreducible objects there, but we now deal with only
symmetric bi-unitary connections, so we have 10 irreducible
objects corresponding to the even vertices.  (Again, it is easy
to have ``symmetric'' ones using the horizonal basic construction.
See Fig. 42 in \cite{BEK2} from a viewpoint of $\alpha$-induction
\cite{BEK1}.)  In this case, we have an example of a
fusion unitary category of tensors which do not come from a 
family of flat connections.  It is not clear what kind of
mathematical assumptions we need for the setting of
\cite{BMWSHV}, but the case we are interested in most is
when we have a unitary fusion category as in 
\cite[Appendix A]{BMWSHV}.  It is well-known that any
unitary fusion category is realized with a family of
flat connections from a (possibly reducible) subfactor as
in Section 2.  (See \cite[Section 12.5]{EK3}, for example.)
So we have the following theorem.

\begin{theorem}
Suppose tensors satisfying 
Bultinck-Mari\"ena-Williamson-\c Sahino\u glu-Haegemana-Verstraete
\cite{BMWSHV} give a unitary fusion category through the
matrix product operator algebra.
Then there exists a subfactor that produces the fusion
category through a family of flat symmetric bi-unitary
connections, and the tube algebra arising from such connections
and the anyon algebra arising from the tensors are isomorphic.
\end{theorem}

Note that a problem of finding (a family of) flat symmetric
bi-unitary connections for
a unitary fusion category given by a family of
non-flat symmetric bi-unitary connections is known as 
that of computing the ``flat part'' of a 
non-flat symmetric bi-unitary connection.

That is, if we have a finite family of
non-flat symmetric bi-unitary connections closed
under the product of such connections, computing the
flat part gives a new family of 
flat symmetric bi-unitary connections. The zipper 
condition holds both for 
non-flat and flat symmetric bi-unitary connections,
but the latter give a canonical form of tensors
and are more appropriate for actual computations
of tensors.

\end{document}